\documentclass[manuscript]{acmart}
\AtBeginDocument{%
  \providecommand\BibTeX{{%
    \normalfont B\kern-0.5em{\scshape i\kern-0.25em b}\kern-0.8em\TeX}}}

\setcopyright{acmcopyright}
\copyrightyear{2021}
\acmYear{2021}
\setcopyright{rightsretained}
\acmConference[RecSys '21]{Fifteenth ACM Conference on Recommender Systems}{September 27-October 1, 2021}{Amsterdam, Netherlands}
\acmBooktitle{Fifteenth ACM Conference on Recommender Systems (RecSys '21), September 27-October 1, 2021, Amsterdam, Netherlands}\acmDOI{10.1145/3460231.3478886}
\acmISBN{978-1-4503-8458-2/21/09}

\usepackage{graphicx}
\usepackage{subfig}

\begin{document}

\title{Multi-Step Critiquing User Interface for Recommender Systems}

\author{Diana Petrescu}\authornote{Both authors contributed equally to this research.}
\author{Diego Antognini}\authornotemark[1]
\author{Boi Faltings}
\affiliation{%
  \institution{École Polytechnique Fédérale de Lausanne}
  \city{Lausanne}
  \country{Switzerland}
}

%% \renewcommand{\shortauthors}{Trovato and Tobin, et al.}

\begin{abstract}
Recommendations with personalized explanations have been shown to increase user trust and perceived quality and help users make better decisions. Moreover, such explanations allow users to provide feedback by critiquing them. Several algorithms for recommender systems with multi-step critiquing have therefore been developed. However, providing a user-friendly interface based on personalized explanations and critiquing has not been addressed in the last decade. In this paper, we introduce four different web interfaces (available under \url{https://lia.epfl.ch/critiquing/}) helping users making decisions and finding their ideal item. We have chosen the hotel recommendation domain as a use case even though our approach is trivially adaptable for other domains. Moreover, our system is model-agnostic (for both recommender systems and critiquing models) allowing a great flexibility and further extensions. Our interfaces are above all a useful tool to help research in recommendation with critiquing. They allow to test such systems on a real use case and also to highlight some limitations of these approaches to find solutions to overcome them.
\end{abstract}

%\begin{CCSXML}
%<ccs2012>
%<concept>
%<concept_id>10003120.10003121.10003129.10011756</concept_id>
%<concept_desc>Human-centered computing~User interface %programming</concept_desc>
%<concept_significance>500</concept_significance>
%</concept>
%<concept>
%<concept_id>10002951.10003227.10003241</concept_id>
%<concept_desc>Information systems~Decision support systems</concept_desc>
%<concept_significance>500</concept_significance>
%</concept>
%</ccs2012>
%\end{CCSXML}

%\ccsdesc[500]{Human-centered computing~User interface programming}
%\ccsdesc[500]{Information systems~Decision support systems}



\maketitle

\section{Introduction}
Recommender systems are intensively researched and widely used nowadays. However, modern recommender systems are often seen as a black box from the end user's point of view. Some studies have shown that user satisfaction with recommender systems does not always correlate with the accuracy they achieve ~\cite{AccuracyHurt} and that providing an explanation to the user as to why certain items are recommended would increase the overall system transparency ~\cite{Transparency, ExplainingRecommendation} and trustworthiness~\cite{Kunkel2018TrustrelatedEO, Zhang2018ExploringEE} and help users make better decisions~\cite{bellini2018knowledge, chang2016crowd}.

Another great advantage of these explanations is that by understanding what generated the recommendations, users can give feedback to the system by critiquing them if they are not satisfied. Critiquing can be viewed as a conversational recommendation method that incorporates user preference feedback and incrementally updates recommendations based on the user's stated preferences~\cite{CritiquingBasedRecommendersPu}. Moreover, it is most useful when applied in multiple steps leading to the development of multi-step critiquing algorithms \cite{multistepcritiquingNeal, SmartClientPuFaltings} that were proven to be better than other methods~\cite{CompoundCritiquing}. 

Research on critiquing has led to the development of methods where the attributes are mined from past reviews. These generated attributes can be in the form of keyphrases used as explanations and allow users to interact with them \cite{antognini2020interacting, luo2020, wu2019deep}. In these approaches, users can add or remove certain aspects that they like or, respectively, dislike or that they consider as significant for the recommendation or not. Moreover, it has been shown that highly personalized natural language justifications increase the perceived recommendation quality and trustworthiness~\cite{kunkel2019let, Kunkel2018TrustrelatedEO, ExplainingRecommendation, chang2016crowd}. Recent works ~\cite{antognini2020interacting,chen2020towards} have proposed approaches to generate these kinds of justification in addition to the keyphrase~explanations.

In this paper, we present several user-friendly web interfaces based on existing explainable recommender systems and critiquing mechanisms~\cite{antognini2020interacting, luo2020}. These interfaces assist users in the decision-making process and help them to better understand the recommended items by providing explanations in the form of actionable keywords and personalized justifications. By critiquing the explanations, the users interact with the system to refine and improve the recommendations by explicitly expressing their preferences. Providing such interfaces has not been addressed in the last decade~\cite{CritiquingBasedRecommendersPu, HybridChen}. More precisely, we have chosen hotel recommendations as a case study as it is a familiar domain for most people. However, our implementation does not depend on a particular dataset and is trivially adaptable for other domains.

In \cite{antognini2020interacting, luo2020}, the evaluation of critiquing is simulated without the participation of actual end users. Our interfaces provide the opportunity to experiment recommendations with critiquing based on neural natural language processing techniques on a real use case. Moreover, they can help to highlight the limitations of these systems to find solutions to overcome them; they are not intended to be a finished usable product for end consumers.

\section{System Overview and Implementation} \label{sysoverview}
\begin{figure}[!tbp]
  \centering
  \subfloat[System interaction.]{\includegraphics[width=0.49\textwidth]{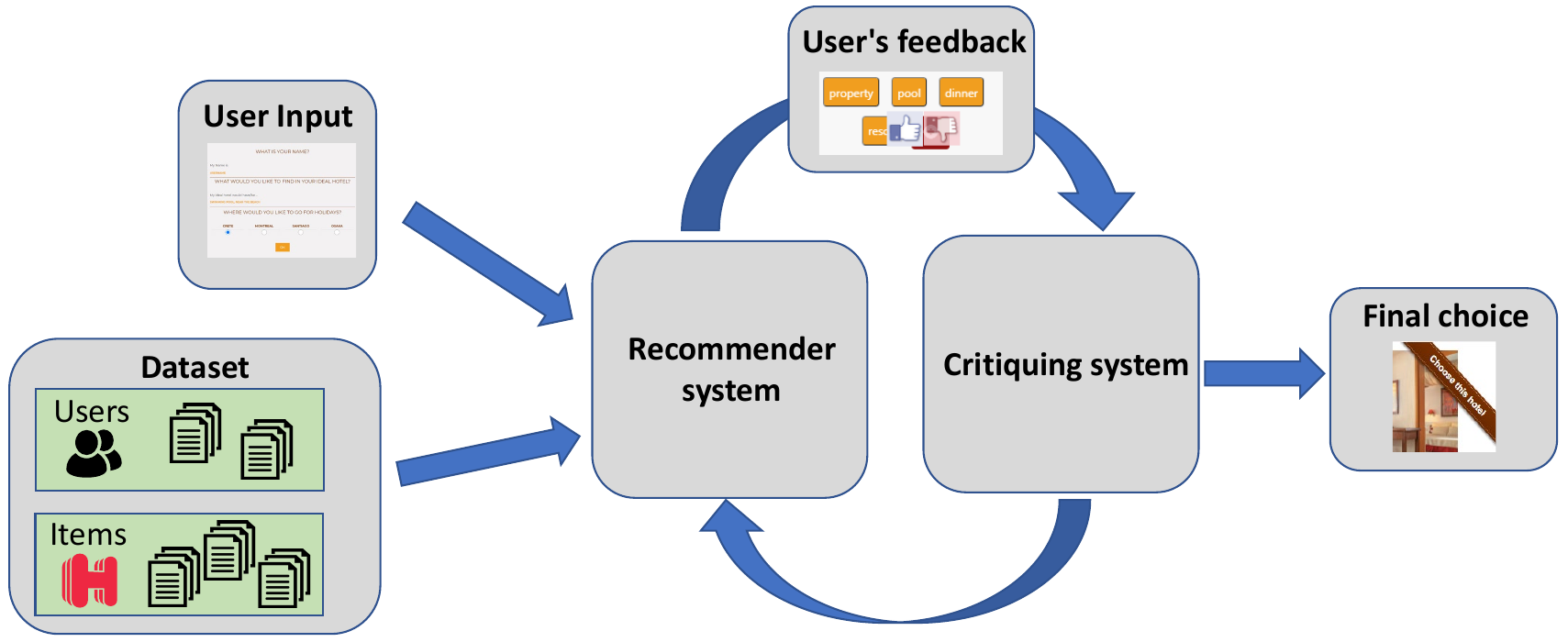}
  \label{fig:systemoverview}}
  \hspace{2mm} %%%%%\hfill
  \subfloat[User interaction.]{\includegraphics[width=0.48\textwidth]{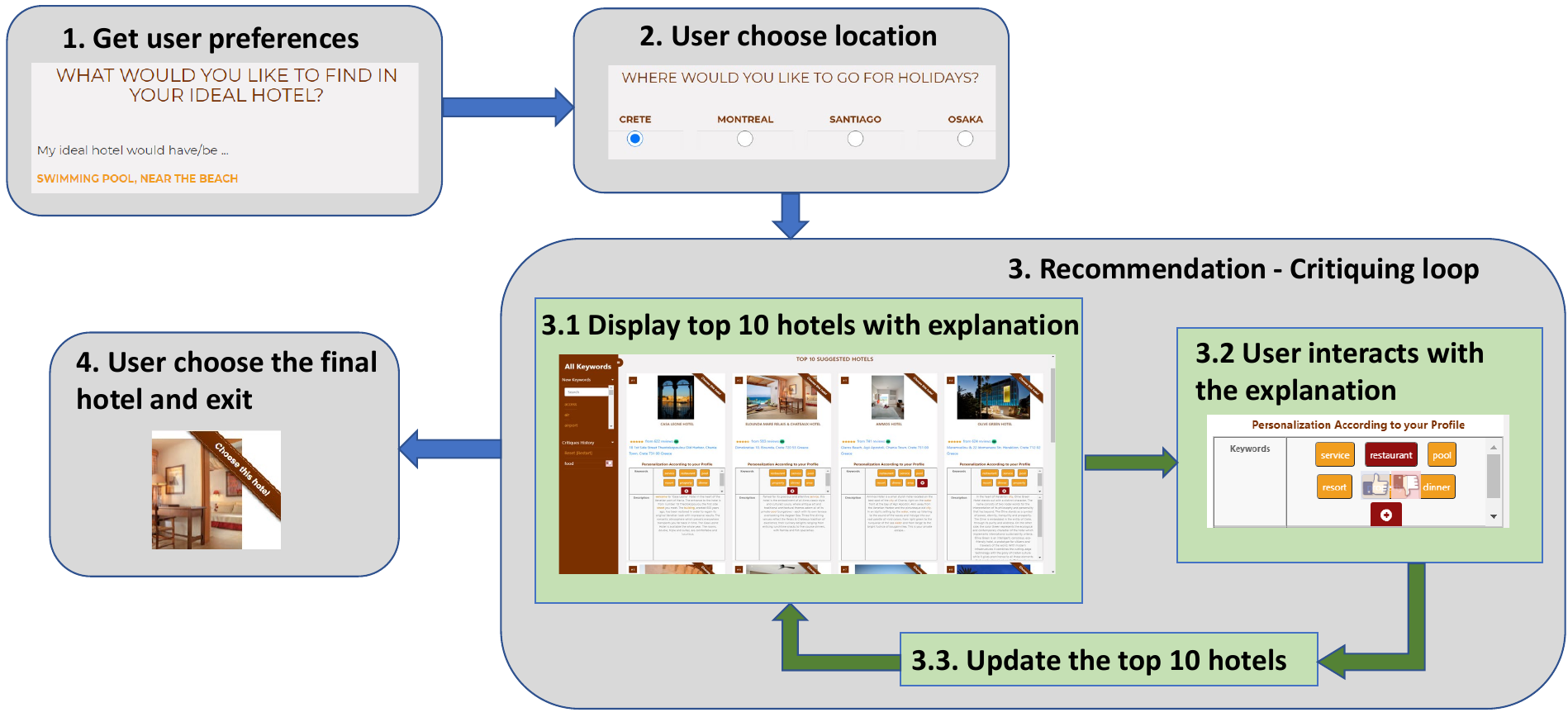}
  \label{fig:useroverview}}
  \caption{Interactions overview.}
\end{figure}

The system overview is shown in Figure~\ref{fig:systemoverview}. To illustrate the flexibility of our system, we employ the HotelRec dataset~\cite{antognini-faltings:2020:LREC1}, which contains 50 million reviews from \textit{TripAdvisor}, and automatically extracted 90 keyphrases to reflect hotel features (similar to \textit{TripAdvisor} or \textit{Booking}). For more details about the preprocessing, we refer the reader to~\cite{antognini2020interacting,antognini2019multi,luo2020}.

Moreover, our system is model-agnostic both for the recommender model and the critiquing mechanism. In case the recommender model does not handle the cold-start problem, we ask the users to provide an input to get their preferences and to define their initial profiles. We then match the query with an existing user in the database whose profile is most similar to the preferences formulated (according to a TF-IDF vector space model computed on user reviews). This enables more flexibility in the used recommender model.

In this demo, we experiment with two recommender systems with critiquing mechanisms: T-RECS~\cite{antognini2020interacting} and CE-VAE~\cite{luo2020}. Both predict a set of keywords from the 90 keyphrases extracted during preprocessing, that best describes~the user profile, alongside the recommended items. These keywords are therefore used as an explanation for the recommendation and can then be critiqued. However, these two models have their own specificities that we detail in~Section~\ref{sec_int}.

The interaction with the interface from the user's point of view is shown in Figure~\ref{fig:useroverview}. Users are first asked to enter, in text form, what they would like for their ideal hotel. Then, users choose one of the four proposed destinations, where each contains 25 to 45 hotels. Next, users enter the \textit{Recommendation - Critiquing loop}. The interface initially displays only the top 10 items recommended based on initial user input, along with the generated explanation. The users have the possibility to interact with the explanations by critiquing them. After each critique, the system updates the recommendation for the users. This process continues until users are satisfied and cease to provide additional critiques. At this point, this iterative process terminates as we consider that users have found their ideal hotel.

\section{Interface Details}
\label{sec_int}
\begin{figure}[!tbp]
  \centering
  \subfloat[Int. A: Menu.]{\quad\quad\includegraphics[height=4.1cm]{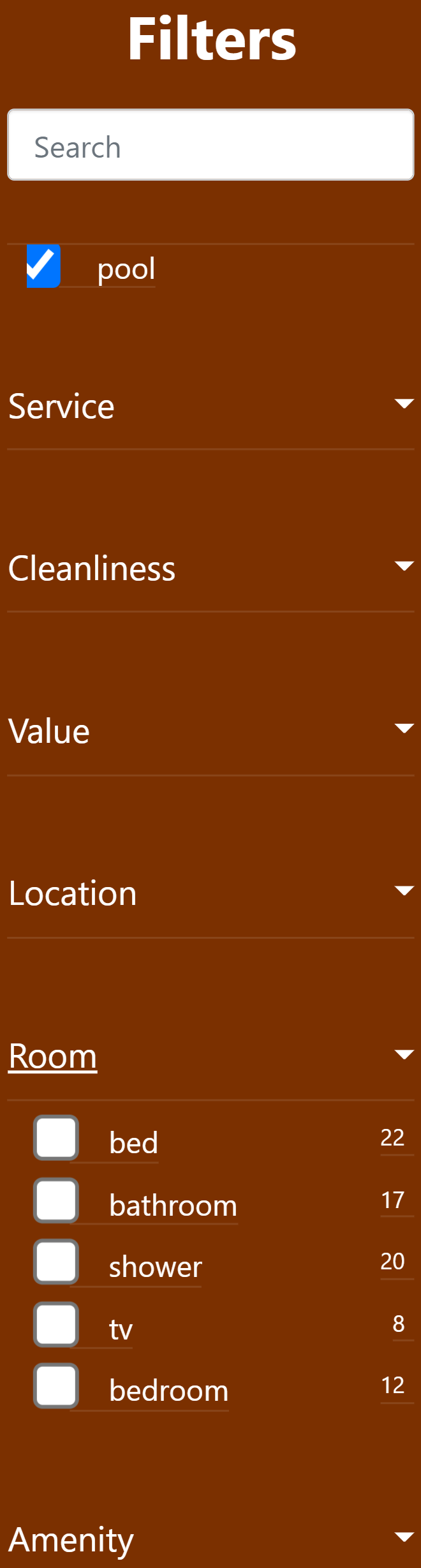}
  \label{fig:menuA}\quad\quad}
  \hspace{1mm}
  \subfloat[Int. B/C/D: Menu.]{\quad\quad\includegraphics[height=4.1cm]{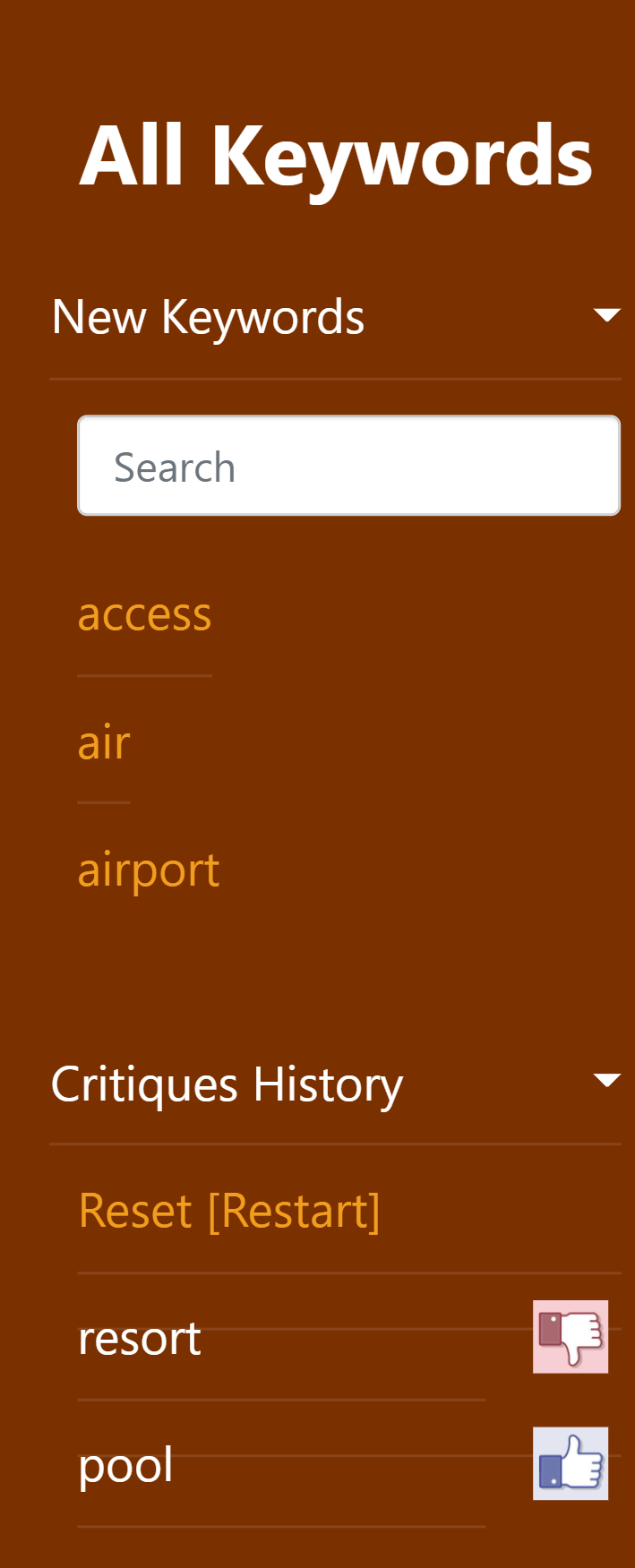}
  \label{fig:menuBCD}\quad\quad}
  \hspace{1mm}
  \subfloat[Int. B: Explanation shared by all items for CE-VAE with descriptions.]{\quad\quad\quad\quad \includegraphics[height=4.1cm]{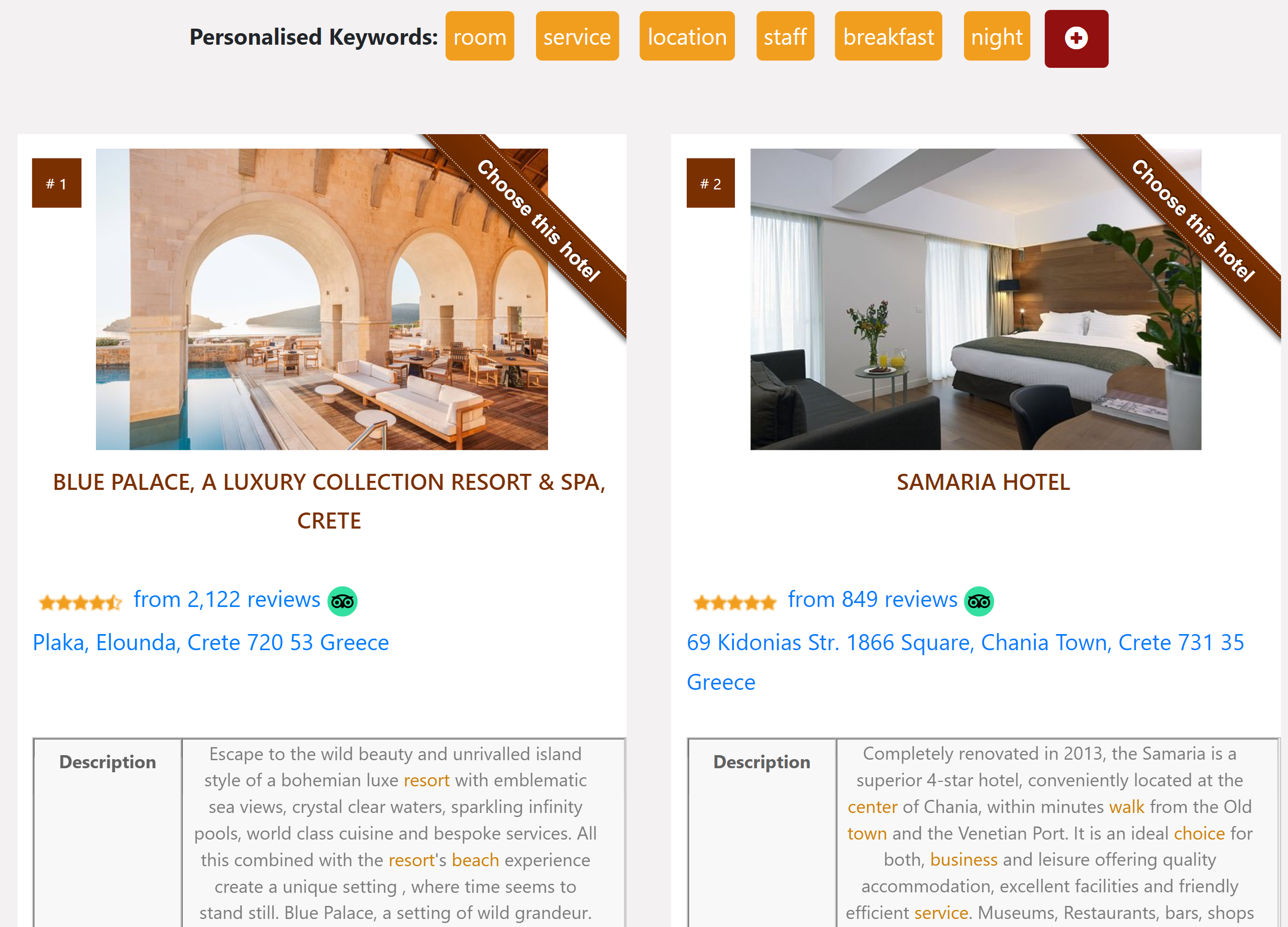}
  \label{fig:interfaceB}\quad\quad\quad\quad}
  \hspace{1mm}
  \\
  \subfloat[Int. C: Explanation per item for TRECS with descriptions.]{\includegraphics[width=0.45\textwidth]{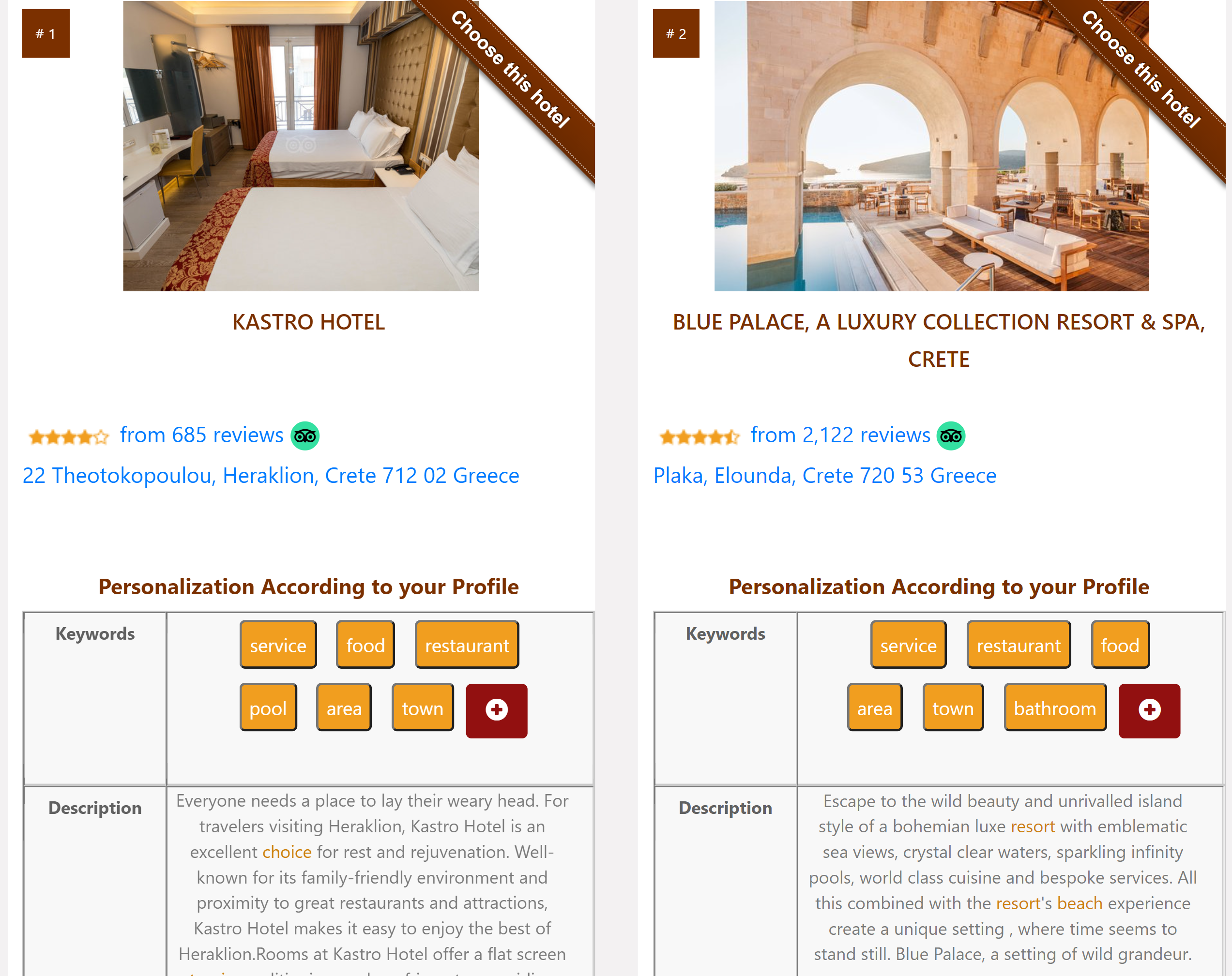}
  \label{fig:interfaceC}}
  \hspace{4mm}
  \subfloat[Int. D: Explanation per item for TRECS with justifications.]{\includegraphics[width=0.48\textwidth]{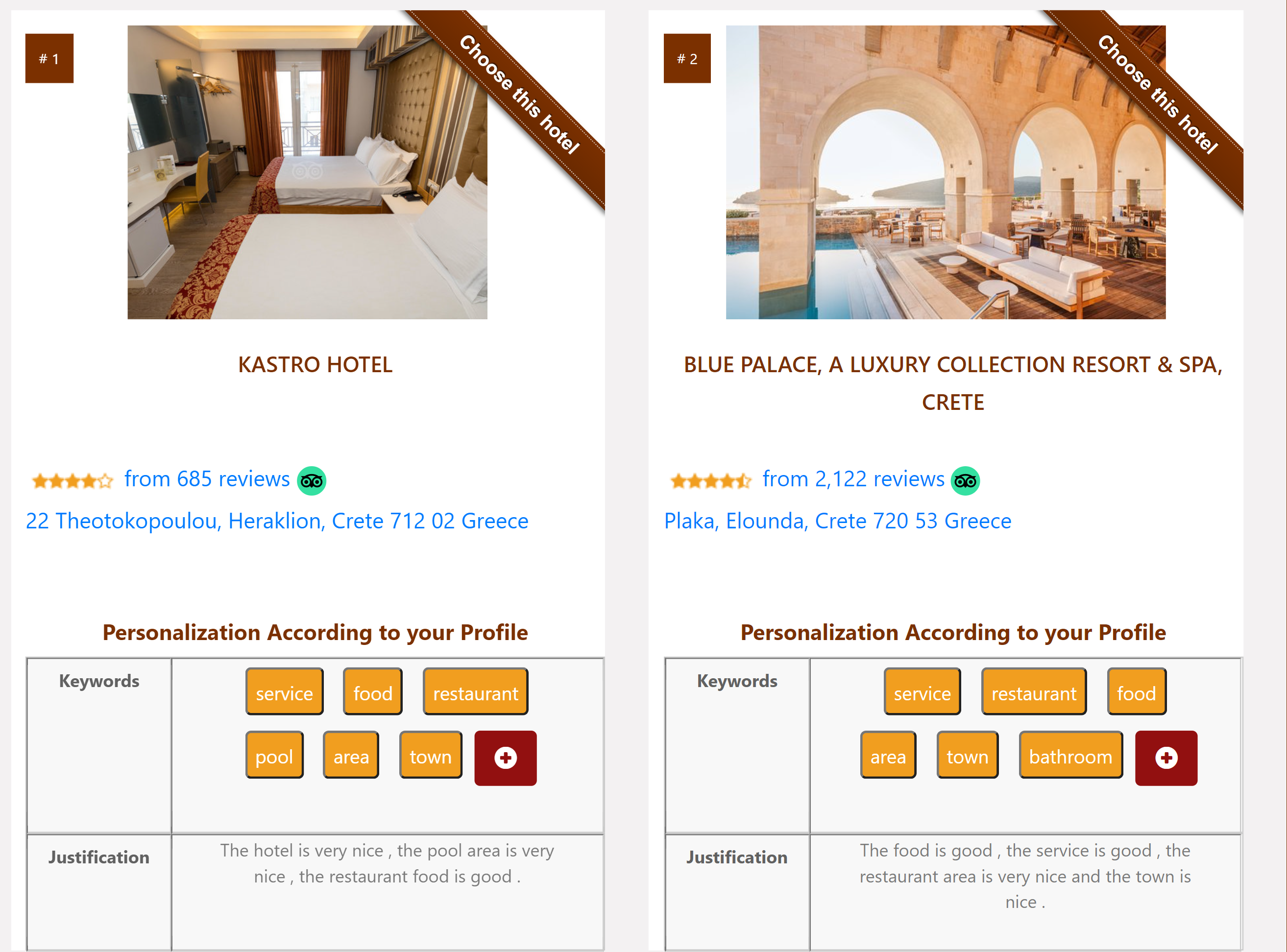}
  \label{fig:interfaceD}}
  \caption{Design details of the four interfaces Int. A/B/C/D.}
  \label{fig:interfaces}
\end{figure}
In fact, users can interact with four different interfaces, some aspects of which are detailed below and shown in Figure~\ref{fig:interfaces}.

\textit{\textbf{Interface A: Static}}.
We use as a reference point of user satisfaction a static interface inspired by current hotel recommender platforms and allowing the user to filter hotels according to the keyword features (which are the same as the extracted keyphrases during preprocessing). To make interaction easier, the keywords are grouped in 6~categories chosen by hand (see Figure~\ref{fig:menuA}). In addition, to provide more information about the hotels and help users make a better choice, the interface also displays the hotel's description scraped from \textit{TripAdvisor}.

\textit{\textbf{Interface B: CE-VAE with shared keyphrases}}.
As explained in Section~\ref{sysoverview}, the recommender system predicts a set of keywords that best describes the user profile alongside the recommended items. For \textit{CE-VAE}, users can only interact with the keywords on a global level to express an explicit disagreement: all items share the same explanation. The top 6~keywords associated with the user profile are displayed at the top of the web page (see Figure~\ref{fig:interfaceB}).
Similar to \textit{Interface~A}, the hotel descriptions are also displayed. However, the system's keywords are highlighted in the descriptions, and it allows the users to also interact with these descriptions by critiquing them.

\textit{\textbf{Interface C: T-RECS with fine-grained keyphrases}}. 
In contrast to \textit{CE-VAE}, \textit{T-RECS} generates more fine-grained explanations by inferring personalized keywords for each user-item pair (see Figure~\ref{fig:interfaceC}). Moreover, users can not only critique a keyword to reflect a disagreement but also highlight it as an important desired feature. As for \textit{Interface B}, the hotel descriptions are also displayed and enable critiquing interactions.

\textit{\textbf{Interface D: T-RECS with fine-grained keyphrases and generated justifications}}. This interface is similar to \textit{Interface C} but, instead of displaying the general scrapped description for each hotel, it generates a personalized natural language justification conditioned on the keyword explanation. This justification, which is therefore based on user profiles, allows them to understand why those hotels are recommended. An example is shown in Figure~\ref{fig:interfaceD}.

In addition, for \textit{Interfaces B, C}, and \textit{D}, we give the possibility to search for additional keywords to critique (see Figure~\ref{fig:menuBCD}) than those predicted as relevant by the system. Finally, after experimenting with the four interfaces, users access the entire hotel catalog to determine if the different approaches helped them finding the hotel that best suits~them.

\section{Discussion}
With the recent progress of natural language generation, models are now capable of generating syntactically and grammatically correct sentences~\cite{radford2019language}. However, we observe that the generated personalized justifications quite often lack diversity~\cite{ni-etal-2019-justifying}. Indeed, the justifications of the best items contain repeating parts of sentences. This makes it more difficult for users to choose their preferred item because the differences between the justifications of the recommended items are not emphasized. In addition, as is often the case in natural language applications, hallucinations can occur~\cite{rohrbach-etal-2018-object,holtzman2019curious}.

Moreover, by experimenting with the interfaces, other limitations of the natural language processing methods used can be observed. We can take as an example the generated keywords. They are obtained by automated systems that generate them based on data mining techniques or machine learning models. However, meaningful keywords for the recommendation do not necessarily imply that they are useful for the end user. On the one hand, the meanings of individual keywords are not always clear for the end users, especially if they are not familiar with the recommendation domain. On the other hand, it is not always obvious what critiquing a keyword means. Indeed, the keyword “pool” indicates the presence or the absence of a pool. However, for the keyword “breakfast”, it is not clear how breakfast is interpreted by the system as most hotels propose breakfast options. Post-processing to filter vague keywords, or the usage of n-grams could improve the understanding of the keywords but will solve the problem only partially.

\section{Conclusion}
Nowadays, recommender systems are widely studied and used. We have designed several web interfaces that can be easily connected to different datasets, recommender models, and critiquing mechanisms. In addition to the usual recommendations, they can provide users with keyword explanations and personalized natural language justifications. Moreover, users can interact with these keywords in different ways by critiquing them. We have focused on the hotel recommendation domain. This gives the opportunity to experiment recommendations with critiquing based on neural natural language processing on a real, most likely familiar use case. The interfaces are also a useful tool to highlight the limitations of these systems and to find solutions to overcome them.

%%%%% \begin{acks}
%%%%% \end{acks}

\bibliographystyle{ACM-Reference-Format}
\bibliography{m_biblio}

%\appendix

\end{document}